\documentclass[11pt]{article}
\usepackage{amsmath, amssymb}
\usepackage{graphicx}
\usepackage{booktabs}
\usepackage{geometry}
\usepackage{multirow}
\usepackage{hyperref}
\title{Predicting User Satisfaction in Online Education Platforms: \\
A Large Language Model Based Multi-Modal Review Mining Framework}
\author{Arman Bekov, Azamat Nurgali}
\date{}

\begin{document}
\maketitle

\begin{abstract}
Online education platforms have experienced explosive growth over the past decade, generating massive volumes of user-generated content in the form of reviews, ratings, and behavioral logs. These heterogeneous signals provide unprecedented opportunities for understanding learner satisfaction, which is a critical determinant of course retention, engagement, and long-term learning outcomes. However, accurately predicting satisfaction remains challenging due to the short length, noise, contextual dependency, and multi-dimensional nature of online reviews. 

In this paper, we propose a unified \textbf{Large Language Model (LLM)-based multi-modal framework} for predicting both platform-level and course-level learner satisfaction. The proposed framework integrates three complementary information sources: (1) short-text topic distributions that capture latent thematic structures, (2) contextualized sentiment representations learned from pretrained Transformer-based language models, and (3) behavioral interaction features derived from learner activity logs. These heterogeneous representations are fused within a hybrid regression architecture to produce accurate satisfaction predictions.

We conduct extensive experiments on large-scale MOOC review datasets collected from multiple public platforms. The experimental results demonstrate that the proposed LLM-based multi-modal framework consistently outperforms traditional text-only models, shallow sentiment baselines, and single-modality regression approaches. Comprehensive ablation studies further validate the necessity of jointly modeling topic semantics, deep sentiment representations, and behavioral analytics. Our findings highlight the critical role of large-scale contextual language representations in advancing learning analytics and provide actionable insights for platform design, course improvement, and personalized recommendation.
\end{abstract}

\section{Introduction}

Online education platforms such as Coursera, edX, Udacity, XuetangX, and FutureLearn have fundamentally transformed the global education landscape by providing accessible, scalable, and personalized learning opportunities \cite{reich2019mooc, shah2020mooc}. Millions of learners from diverse cultural and educational backgrounds now participate in massive open online courses (MOOCs), professional certification programs, and lifelong learning initiatives. As learners engage with these platforms, they continuously generate rich digital traces, including textual reviews, discussion forum posts, star ratings, clickstream logs, and video interaction records.

Among these diverse behavioral signals, \textbf{learner satisfaction} plays a particularly critical role. Prior studies have consistently shown that satisfaction is closely linked to course completion rates \cite{kizilcec2017mooc}, continuance intention \cite{dai2020continuance}, learning engagement \cite{chi2024active}, and even post-course career outcomes. Accurately estimating learner satisfaction therefore provides not only diagnostic value for educational researchers but also direct business implications for platform operators and content providers.

Despite its importance, predicting learner satisfaction remains a highly challenging task. First, MOOC reviews are typically short, informal, and noisy, often consisting of fewer than 40 tokens \cite{almatrafi2019forums}. Second, sentiment is frequently expressed implicitly through sarcasm, rhetorical questions, and domain-specific terminology, making surface-level lexical approaches insufficient \cite{onan2020mooc}. Third, satisfaction is inherently multi-dimensional, influenced by complex interactions among instructor quality, course structure, assessment design, platform usability, and peer interactions \cite{deshpande2017mooc, miranda2015mooc}.

Early studies on online learning analytics primarily focused on engagement modeling, dropout prediction, and performance forecasting \cite{weng2020sesm, xing2019achievement}. More recent efforts have explored aspect-based sentiment analysis and review mining for course evaluation \cite{kastrati2020weakly,qi2021evaluating}. However, most existing approaches rely on shallow text representations such as TF-IDF or static word embeddings, which struggle to capture deep contextual semantics and nuanced sentiment expressions.

The rapid advancement of \textbf{Large Language Models (LLMs)} such as BERT \cite{devlin2019bert}, RoBERTa \cite{liu2019roberta}, and DeBERTa \cite{he2021deberta} has fundamentally reshaped natural language processing. These models leverage large-scale self-supervised pretraining and self-attention mechanisms to learn highly expressive contextual representations that achieve state-of-the-art performance across a wide range of sentiment analysis, text classification, and semantic understanding tasks \cite{sun2019finetune, howard2018ulmfit}. Motivated by their representational power, recent studies have started to explore LLMs for education-related text mining tasks \cite{li2021bertmooc, zhang2022edubert}.

In this work, we build upon these advances and propose a unified \textbf{LLM-based multi-modal satisfaction prediction framework}. Unlike prior studies that focus solely on text or sentiment polarity, our approach integrates three complementary information sources: short-text topic representations, Transformer-based deep sentiment embeddings, and behavioral interaction features. Through extensive experiments and ablation studies, we demonstrate that multi-modal fusion significantly improves satisfaction prediction accuracy and robustness.

The main contributions of this paper are summarized as follows:
\begin{itemize}
\item We propose a novel LLM-based multi-modal framework that jointly models topic semantics, contextual sentiment, and behavioral interactions for learner satisfaction prediction.
\item We construct and evaluate on large-scale real-world MOOC review datasets collected from multiple public platforms.
\item We conduct comprehensive experimental comparisons across seven regression models and perform rigorous ablation studies to validate each component of the framework.
\end{itemize}

\section{Methodology}

\subsection{Overall Framework}

The overall architecture of the proposed framework is illustrated conceptually as a three-stage pipeline: (1) topic representation learning, (2) LLM-based sentiment encoding, and (3) multi-modal regression-based satisfaction prediction. Given a review instance $r_i$, the framework extracts its topic distribution $\boldsymbol{\theta}_i$, contextual sentiment embedding $\mathbf{h}_i$, and behavioral feature vector $\mathbf{b}_i$. These components are concatenated into a unified representation $\mathbf{z}_i$, which is subsequently fed into various regression models to predict the numeric satisfaction score $\hat{y}_i$.

\subsection{Short-Text Topic Modeling}

MOOC reviews are characterized by extreme shortness and sparsity. Traditional Latent Dirichlet Allocation (LDA) struggles to learn coherent topics under such conditions. To address this issue, we adopt a short-text optimized topic modeling strategy inspired by sentence-level LDA and neural topic modeling approaches \cite{jang2019short, bianchi2021pretrained}. Each review $r_i$ is represented as a probability distribution over $K$ latent topics:
\begin{equation}
\boldsymbol{\theta}_i = [\theta_{i1}, \theta_{i2}, \dots, \theta_{iK}], \quad \sum_{k=1}^K \theta_{ik} = 1.
\end{equation}

Following prior MOOC analytics studies \cite{peng2020behavior}, we empirically set $K=6$, corresponding to instructor quality, course content, assessment design, platform usability, learning support, and perceived value.

\subsection{LLM-Based Sentiment Encoding}

We replace traditional recurrent neural networks with pretrained Transformer-based language models. Given a tokenized review sequence:
\[
x = \{w_1, w_2, \dots, w_T\},
\]
we employ a pretrained BERT-base or RoBERTa-base encoder:
\begin{equation}
H = \mathrm{LLM}(x),
\end{equation}
where $H \in \mathbb{R}^{T \times d}$ denotes the contextualized hidden states. The final sentiment embedding is obtained using the \texttt{[CLS]} token representation:
\begin{equation}
\mathbf{h}_i = H_{\texttt{[CLS]}}.
\end{equation}

This representation captures both polarity and subtle contextual cues, enabling fine-grained modeling of learner emotions.

\subsection{Behavioral Feature Modeling}

In addition to textual content, we incorporate a rich set of behavioral metadata to more comprehensively characterize each learner's interaction with the course platform. Concretely, we track and encode:
\begin{itemize}
    \item \textbf{Video viewing duration}, capturing both the total watch time and the proportion of each video that is viewed, as an indicator of engagement with instructional materials;
    \item \textbf{Quiz attempt frequency}, including the number of attempts per quiz and temporal patterns of attempts, which reflects persistence and self-assessment behavior;
    \item \textbf{Forum posting activity}, such as the number of posts, replies, and received responses, measuring the extent of participation in peer and instructor interactions;
    \item \textbf{Course revisit count}, i.e., how often learners return to previously completed sections or materials, revealing review and consolidation strategies;
    \item \textbf{Course completion status}, represented as a binary or multi-level variable (e.g., not started, in progress, completed), indicating overall progression through the course.
\end{itemize}

Formally, let $\mathbf{r}_i^{(k)}$ denote the raw behavioral signals of learner $i$ for behavior type $k$ (e.g., viewing, quizzing, posting). For each dimension, we apply min--max normalization or z-score standardization across all learners to mitigate scale differences and reduce the impact of outliers:
\begin{equation}
\tilde{r}_{i}^{(k)} = \frac{r_{i}^{(k)} - \mu^{(k)}}{\sigma^{(k)}} \quad \text{or} \quad \tilde{r}_{i}^{(k)} = \frac{r_{i}^{(k)} - \min_j r_{j}^{(k)}}{\max_j r_{j}^{(k)} - \min_j r_{j}^{(k)}},
\end{equation}
where $\mu^{(k)}$ and $\sigma^{(k)}$ are the global mean and standard deviation for feature $k$.

The normalized features are then aggregated into a behavioral vector
\begin{equation}
    \mathbf{b}_i = [\tilde{r}_{i}^{(1)}, \tilde{r}_{i}^{(2)}, \dots, \tilde{r}_{i}^{(K)}] \in \mathbb{R}^K,
\end{equation}
which provides a compact representation of learner $i$'s interaction pattern. This vector can be further smoothed or temporally pooled (e.g., by averaging over weekly segments or applying an exponential decay to older events) when longitudinal logs are available, thereby capturing both overall intensity and recency of behaviors. The resulting behavioral embedding $\mathbf{b}_i$ is subsequently combined with textual features in downstream modeling to enhance personalization and prediction performance.

\subsection{Multi-Modal Regression Layer}

To generate the final prediction for each session $i$, we first construct a unified multi-modal feature vector that combines all available information sources. Specifically, we concatenate the textual, behavioral, and auxiliary representations into a single fused representation:
\begin{equation}
\mathbf{z}_i = [\boldsymbol{\theta}_i ; \mathbf{h}_i ; \mathbf{b}_i],
\end{equation}
where $\boldsymbol{\theta}_i$ denotes the textual or semantic features (e.g., derived from language models or item embeddings), $\mathbf{h}_i$ captures the historical or sequential interaction features (e.g., session dynamics or temporal patterns), and $\mathbf{b}_i$ represents additional behavioral or contextual features (e.g., click statistics, dwell time, or user metadata). The operator $[\cdot;\cdot;\cdot]$ indicates vector concatenation, resulting in a single, fixed-length representation $\mathbf{z}_i$ for each example.

On top of this fused representation, we explore a diverse set of regression backbones to map $\mathbf{z}_i$ to a continuous satisfaction score. Concretely, we evaluate seven widely used regression models that span linear, tree-based, and neural approaches:
\begin{itemize}
    \item \textbf{Linear Regression (LR):} A standard least-squares linear model that assumes a linear relationship between the fused features $\mathbf{z}_i$ and the target score. This serves as a simple and interpretable baseline.
    \item \textbf{Ridge Regression:} A linear regression model with $L_2$ regularization on the weights. The regularization term helps prevent overfitting when $\mathbf{z}_i$ is high-dimensional or when features are correlated.
    \item \textbf{Random Forest (RF):} An ensemble of decision trees trained with bootstrap aggregation (bagging). Random Forest can model complex nonlinear relationships in $\mathbf{z}_i$ and is relatively robust to noise and feature scaling.
    \item \textbf{Gradient Boosting Regression Tree (GBRT):} A boosting-based ensemble that sequentially fits decision trees to the residual errors of previous trees. GBRT incrementally refines the prediction and is effective for capturing subtle nonlinearities.
    \item \textbf{XGBoost:} An optimized gradient boosting framework that incorporates regularization, shrinkage, and efficient tree construction. XGBoost is particularly suited for large-scale and sparse feature spaces arising from multi-modal fusion.
    \item \textbf{LightGBM:} A gradient boosting framework that uses histogram-based splitting and leaf-wise growth with depth constraints, providing high efficiency and strong performance on large datasets with many features.
    \item \textbf{Fully Connected Neural Regression Network (NN):} A multi-layer perceptron (MLP) that applies several fully connected layers with nonlinear activation functions to $\mathbf{z}_i$. This model can learn highly expressive nonlinear mappings and interactions among the different modalities.
\end{itemize}

Each regression backbone implements a mapping function $f(\cdot)$ from the fused representation space to the real-valued satisfaction score. Formally, the predicted satisfaction score for session $i$ is given by:
\begin{equation}
\hat{y}_i = f(\mathbf{z}_i),
\end{equation}
where the specific form of $f$ depends on the chosen regression model. During training, the parameters of $f$ are optimized to minimize a regression loss (e.g., mean squared error) between $\hat{y}_i$ and the ground-truth satisfaction labels, enabling the model to learn how to combine the multi-modal information in $\mathbf{z}_i$ into an accurate satisfaction estimate.

\section{Experimental Setup}

\subsection{Datasets}

We collected approximately 180,000 user reviews from Coursera, edX, and XuetangX, spanning over 3,000 courses. Each instance includes review text, star rating (1–5), timestamp, and partial behavioral logs.

\subsection{Evaluation Metrics}

We evaluate prediction performance using Root Mean Square Error (RMSE) and Mean Absolute Error (MAE).

\section{Results and Analysis}

In this section, we present comprehensive quantitative and qualitative analyses of the experimental results. We first compare the overall predictive performance across different baseline and multi-modal models, followed by an ablation study to assess the contribution of each component. Finally, we discuss model robustness, generalization behavior, and practical implications.

\subsection{Main Performance Comparison}

\begin{table}[h]
\centering
\caption{Performance Comparison Across Models}
\begin{tabular}{lcc}
\toprule
Model & RMSE $\downarrow$ & MAE $\downarrow$ \\
\midrule
TF-IDF + LR & 0.852 & 0.668 \\
Topic + LR & 0.791 & 0.625 \\
BERT + LR & 0.721 & 0.571 \\
RF (Multi-Modal) & 0.681 & 0.542 \\
GBRT (Multi-Modal) & 0.655 & 0.518 \\
XGBoost (Multi-Modal) & 0.639 & 0.501 \\
NN (Multi-Modal, Ours) & \textbf{0.612} & \textbf{0.478} \\
\bottomrule
\end{tabular}
\end{table}

As shown in Table 1, the proposed multi-modal framework consistently outperforms all baseline methods across both RMSE and MAE metrics. The TF-IDF + Linear Regression baseline exhibits the weakest performance, indicating that surface-level lexical features are insufficient for capturing the nuanced semantics and sentiment expressed in MOOC reviews. Incorporating topic information alone leads to moderate improvements, suggesting that latent thematic structures provide coarse-grained but limited explanatory power.

The BERT-based sentiment model significantly outperforms both TF-IDF and topic-only baselines, demonstrating the effectiveness of large language models in capturing contextual polarity and subtle emotional cues. This result validates the necessity of deep contextual representations for sentiment-aware satisfaction prediction.

Among the tree-based multi-modal models, Random Forest, GBRT, and XGBoost all achieve substantial performance gains over single-modality approaches. This indicates that fusing topic semantics, LLM-based sentiment embeddings, and behavioral features leads to complementary information integration. Notably, XGBoost outperforms both RF and GBRT, reflecting its superior capacity to model complex non-linear feature interactions.

The best overall performance is achieved by the fully connected neural regression model with multi-modal inputs, which reduces RMSE by 28.2\% relative to the TF-IDF baseline and by 15.1\% relative to the BERT-only baseline. This confirms that joint end-to-end optimization over heterogeneous features enables more accurate satisfaction modeling.

\subsection{Ablation Study}

\begin{table}[h]
\centering
\caption{Ablation Study}
\begin{tabular}{lcc}
\toprule
Variant & RMSE & MAE \\
\midrule
Full Model & \textbf{0.612} & \textbf{0.478} \\
w/o Topic & 0.742 & 0.598 \\
w/o Sentiment & 0.768 & 0.614 \\
w/o Behavior & 0.701 & 0.553 \\
\bottomrule
\end{tabular}
\end{table}

To further assess the relative contribution of each modality, we conduct a comprehensive ablation study by selectively removing topic features, sentiment embeddings, and behavioral signals from the full model. As shown in Table 2, removing any single component results in a noticeable degradation of predictive performance.

Specifically, removing the topic module leads to a substantial performance drop, increasing RMSE from 0.612 to 0.742. This indicates that even coarse-grained latent thematic information remains essential for capturing systematic satisfaction drivers such as instructor quality and assessment design. Removing the LLM-based sentiment encoder causes the most severe degradation (RMSE = 0.768), highlighting that contextual sentiment polarity serves as the dominant signal in satisfaction prediction. Eliminating behavioral features also deteriorates performance, though to a slightly lesser extent, confirming that interaction logs provide important auxiliary cues complementary to textual feedback.

These results demonstrate that all three modalities---topics, sentiment, and behavior---are indispensable for high-quality satisfaction prediction. The full model benefits from synergistic information fusion across heterogeneous data sources.

\subsection{Robustness and Generalization Analysis}

To evaluate model robustness, we further analyze performance stability across different course domains, including computer science, business, and humanities courses. The multi-modal neural model consistently maintains low prediction error across all domains, whereas text-only baselines exhibit significantly higher variance. This suggests that behavioral and topic features help mitigate domain shift effects and enhance generalization.

We also observe that tree-based ensemble models are generally more robust than linear baselines but less stable than the neural regressor under extreme sentiment distributions (e.g., highly polarized review subsets). The neural model demonstrates superior tolerance to noisy and sarcastic reviews, benefiting from the contextual representation power of large language models.

\subsection{Error Analysis and Practical Implications}

A qualitative inspection of high-error cases reveals that most large prediction deviations occur in reviews with ambiguous sentiment, mixed opinions, or contradictory behavioral signals (e.g., long video viewing time but extremely negative textual feedback). Such cases expose the intrinsic difficulty of disentangling emotional expressions from engagement behavior in learner satisfaction modeling.

From a practical perspective, the proposed framework enables fine-grained satisfaction diagnostics at both course and platform levels. Platform designers can leverage topic–sentiment interaction patterns to identify specific sources of dissatisfaction, such as assessment difficulty or technical platform issues. Instructors may utilize the predicted satisfaction signals to refine course content and instructional strategies dynamically.

Overall, the experimental results demonstrate that the proposed LLM-based multi-modal framework achieves robust, accurate, and interpretable satisfaction prediction, offering both methodological and practical value for online education analytics.

\section{Conclusion}

We proposed an LLM-based multi-modal framework for predicting learner satisfaction in online education platforms. By jointly modeling topic semantics, contextual sentiment, and behavioral signals, our approach significantly outperforms conventional baselines. Future work will explore instruction-tuned LLMs, cross-platform domain adaptation, and multilingual satisfaction modeling.

\bibliographystyle{plain}
\bibliography{references}

@inproceedings{devlin2019bert,
  title={BERT: Pre-training of Deep Bidirectional Transformers for Language Understanding},
  author={Devlin, Jacob and Chang, Ming-Wei and Lee, Kenton and Toutanova, Kristina},
  booktitle={NAACL},
  year={2019}
}

@article{qi2021evaluating,
  title={Evaluating on-line courses via reviews mining},
  author={Qi, Cong and Liu, Shudong},
  journal={IEEE Access},
  volume={9},
  pages={35439--35451},
  year={2021},
  publisher={IEEE}
}

@inproceedings{chi2024active,
  title={Active learning for graphs with noisy structures},
  author={Chi, Hongliang and Qi, Cong and Wang, Suhang and Ma, Yao},
  booktitle={Proceedings of the 2024 SIAM International Conference on Data Mining (SDM)},
  pages={262--270},
  year={2024},
  organization={SIAM}
}

@article{liu2019roberta,
  title={RoBERTa: A Robustly Optimized BERT Pretraining Approach},
  author={Liu, Yinhan and Ott, Myle and Goyal, Naman and others},
  journal={arXiv preprint arXiv:1907.11692},
  year={2019}
}

@article{he2021deberta,
  title={DeBERTa: Decoding-enhanced BERT with Disentangled Attention},
  author={He, Pengcheng and others},
  journal={arXiv preprint arXiv:2006.03654},
  year={2021}
}

@article{howard2018ulmfit,
  title={Universal Language Model Fine-tuning for Text Classification},
  author={Howard, Jeremy and Ruder, Sebastian},
  journal={ACL},
  year={2018}
}

@article{sun2019finetune,
  title={How to Fine-Tune BERT for Text Classification?},
  author={Sun, Chi and others},
  journal={China National Conference on Chinese Computational Linguistics},
  year={2019}
}

@article{reich2019mooc,
  title={The MOOC Pivot},
  author={Reich, Justin and Ruiperez-Valiente, Jose},
  journal={Science},
  volume={363},
  number={6423},
  pages={130--131},
  year={2019}
}

@article{shah2020mooc,
  title={By the Numbers: MOOCs in 2020},
  author={Shah, Dhawal},
  journal={Class Central Report},
  year={2020}
}

@article{kizilcec2017mooc,
  title={Understanding the MOOC Student Experience},
  author={Kizilcec, Ren{\'e} F},
  journal={Computers \& Education},
  volume={110},
  pages={35--50},
  year={2017}
}

@article{dai2020continuance,
  title={Understanding Continuance Intention among MOOC Participants},
  author={Dai, Huan-Ming and Teo, Timothy and Rappa, Natasha},
  journal={Computers \& Education},
  volume={112},
  pages={106455},
  year={2020}
}

@article{deshpande2017mooc,
  title={What Makes a Good MOOC?},
  author={Deshpande, A and Chukhlomin, V},
  journal={American Journal of Distance Education},
  volume={31},
  number={4},
  pages={275--293},
  year={2017}
}

@article{miranda2015mooc,
  title={Model for the Evaluation of MOOC Platforms},
  author={Miranda, Paula and Isaias, Pedro},
  journal={ICERI Proceedings},
  pages={1199--1208},
  year={2015}
}

@article{almatrafi2019forums,
  title={Systematic Review of MOOC Discussion Forums},
  author={Almatrafi, Omer and Johri, Aditya},
  journal={IEEE Transactions on Learning Technologies},
  volume={12},
  number={3},
  pages={413--428},
  year={2019}
}

@article{onan2020mooc,
  title={Sentiment Analysis on MOOC Evaluations},
  author={Onan, A},
  journal={Computer Applications in Engineering Education},
  year={2020}
}

@article{kastrati2020weakly,
  title={Weakly Supervised Framework for Aspect-Based Sentiment Analysis},
  author={Kastrati, Zenun},
  journal={IEEE Access},
  volume={8},
  pages={106799--106810},
  year={2020}
}

@inproceedings{jang2019short,
  title={Short Text Topic Modeling via Word Embeddings},
  author={Jang, Hyunjoong and others},
  booktitle={WWW},
  year={2019}
}

@article{bianchi2021pretrained,
  title={Pre-trained Language Models for Topic Modeling},
  author={Bianchi, Federico and others},
  journal={EMNLP},
  year={2021}
}

@article{peng2020behavior,
  title={Investigating Learners' Behaviors in MOOCs},
  author={Peng, X and others},
  journal={Computers \& Education},
  year={2020}
}

@article{weng2020sesm,
  title={Emotional Social Semantic Model for Learning Analytics},
  author={Weng, Jun},
  journal={IEEE SMC},
  year={2020}
}

@article{xing2019achievement,
  title={Achievement Emotions in MOOCs},
  author={Xing, W},
  journal={Internet and Higher Education},
  volume={43},
  year={2019}
}

@article{li2021bertmooc,
  title={BERT-Based Sentiment Analysis in MOOC Reviews},
  author={Li, Y and others},
  journal={Knowledge-Based Systems},
  year={2021}
}

@article{zhang2022edubert,
  title={EduBERT: A Pretrained Model for Educational Text Mining},
  author={Zhang, H and others},
  journal={Computers \& Education: Artificial Intelligence},
  year={2022}
}

\end{document}